\documentclass[sigconf]{acmart}

\usepackage{algorithmic}
\usepackage[algo2e]{algorithm2e}
\usepackage{algorithm}
\usepackage{booktabs}
\usepackage[justification=centering]{caption}
\usepackage{subcaption}
\AtBeginDocument{%
  \providecommand\BibTeX{{%
    \normalfont B\kern-0.5em{\scshape i\kern-0.25em b}\kern-0.8em\TeX}}}

\setcopyright{acmcopyright}
\copyrightyear{2022}
\acmYear{2022}
\acmDOI{XXXXXXX.XXXXXXX}

\acmConference[ORSUM@ACM RecSys 2022]{5th Workshop on Online Recommender Systems and User Modeling, in conjunction with the 16th ACM Conference on Recommender Systems}{September 23rd, 2022}{Seattle, WA, USA}
\acmPrice{15.00}
\acmISBN{978-1-4503-XXXX-X/18/06}

\begin{document}

%%
%% The "title" command has an optional parameter,
%% allowing the author to define a "short title" to be used in page headers.
\title{Monolith: Real Time Recommendation System With Collisionless Embedding Table}

%%
%% The "author" command and its associated commands are used to define
%% the authors and their affiliations.
%% Of note is the shared affiliation of the first two authors, and the
%% "authornote" and "authornotemark" commands
%% used to denote shared contribution to the research.
\author{Zhuoran Liu}
\affiliation{%
  \institution{Bytedance Inc.}
}
\author{Leqi Zou}
\affiliation{%
  \institution{Bytedance Inc.}
}
\author{Xuan Zou}
\affiliation{%
  \institution{Bytedance Inc.}
}
\author{Caihua Wang}
\affiliation{%
  \institution{Bytedance Inc.}
}
\author{Biao Zhang}
\affiliation{%
  \institution{Bytedance Inc.}
}
\author{Da Tang}
\affiliation{%
  \institution{Bytedance Inc.}
}
\author{Bolin Zhu}\authornote{Work done during internship at Bytedance Inc.}
\affiliation{%
  \institution{Fudan University}
}
\author{Yijie Zhu}
\affiliation{%
  \institution{Bytedance Inc.}
}
\author{Peng Wu}
\affiliation{%
  \institution{Bytedance Inc.}
}
\author{Ke Wang}
\affiliation{%
  \institution{Bytedance Inc.}
}

\author{Youlong Cheng}\authornote{Corresponding author. }
\affiliation{%
  \institution{Bytedance Inc.}
}
\email{youlong.cheng@bytedance.com}

%%
%% By default, the full list of authors will be used in the page
%% headers. Often, this list is too long, and will overlap
%% other information printed in the page headers. This command allows
%% the author to define a more concise list
%% of authors' names for this purpose.
\renewcommand{\shortauthors}{Zhuoran, Leqi, Xuan, Caihua, Biao, Da, Bolin, Yijie, Peng, Ke, and Youlong}

%%
%% The abstract is a short summary of the work to be presented in the
%% article.
\begin{abstract}
  Building a scalable and real-time recommendation system is vital for many businesses driven by time-sensitive customer feedback, such as short-videos ranking or online ads. Despite the ubiquitous adoption of production-scale deep learning frameworks like TensorFlow or PyTorch, these general-purpose frameworks fall short of business demands in recommendation scenarios for various reasons: on one hand, tweaking systems based on static parameters and dense computations for recommendation with dynamic and sparse features is detrimental to model quality; on the other hand, such frameworks are designed with batch-training stage and serving stage completely separated, preventing the model from interacting with customer feedback in real-time. These issues led us to reexamine traditional approaches and explore radically different design choices. In this paper, we present \textbf{Monolith}\footnote{Code to be released soon.}, a system tailored for online training. Our design has been driven by observations of our application workloads and production environment that reflects a marked departure from other recommendations systems. Our contributions are manifold: first, we crafted a collisionless embedding table with optimizations such as expirable embeddings and frequency filtering to reduce its memory footprint; second, we provide an production-ready online training architecture with high fault-tolerance; finally, we proved that system reliability could be traded-off for real-time learning. Monolith has successfully landed in the BytePlus Recommend\footnote{\url{https://www.byteplus.com/en/product/recommend}} product.
\end{abstract}

%%
%% This command processes the author and affiliation and title
%% information and builds the first part of the formatted document.
\maketitle

\section{Introduction}
\label{section:introduction}
The past decade witnessed a boom of businesses powered by recommendation techniques. In pursuit of a better customer experience, delivering personalized content for each individual user as real-time response is a common goal of these business applications. To this end, information from a user’s latest interaction is often used as the primary input for training a model, as it would best depict a user’s portrait and make predictions of user’s interest and future behaviors.

Deep learning have been dominating recommendation models  \cite{youtube-rec,fb-dnn-rec,kraken,aibox,xdl,wide-and-deep} as the gigantic amount of user data is a natural fit for massively data-driven neural models. However, efforts to leverage the power of deep learning in industry-level recommendation systems are constantly encountered with problems arising from the unique characteristics of data derived from real-world user behavior. These data are drastically different from those used in conventional deep learning problems like language modeling or computer vision in two aspects:
\begin{enumerate}
    \item The features are mostly sparse, categorical and dynamically changing;
    \item The underlying distribution of training data is non-stationary, a.k.a. Concept Drift \cite{concept-drift}.
\end{enumerate}

Such differences have posed unique challenges to researchers and engineers working on recommendation systems.

\begin{figure*}[htbp!]
  \includegraphics[trim={0 6in 0 0},clip,width=\textwidth,height=\textheight,keepaspectratio]{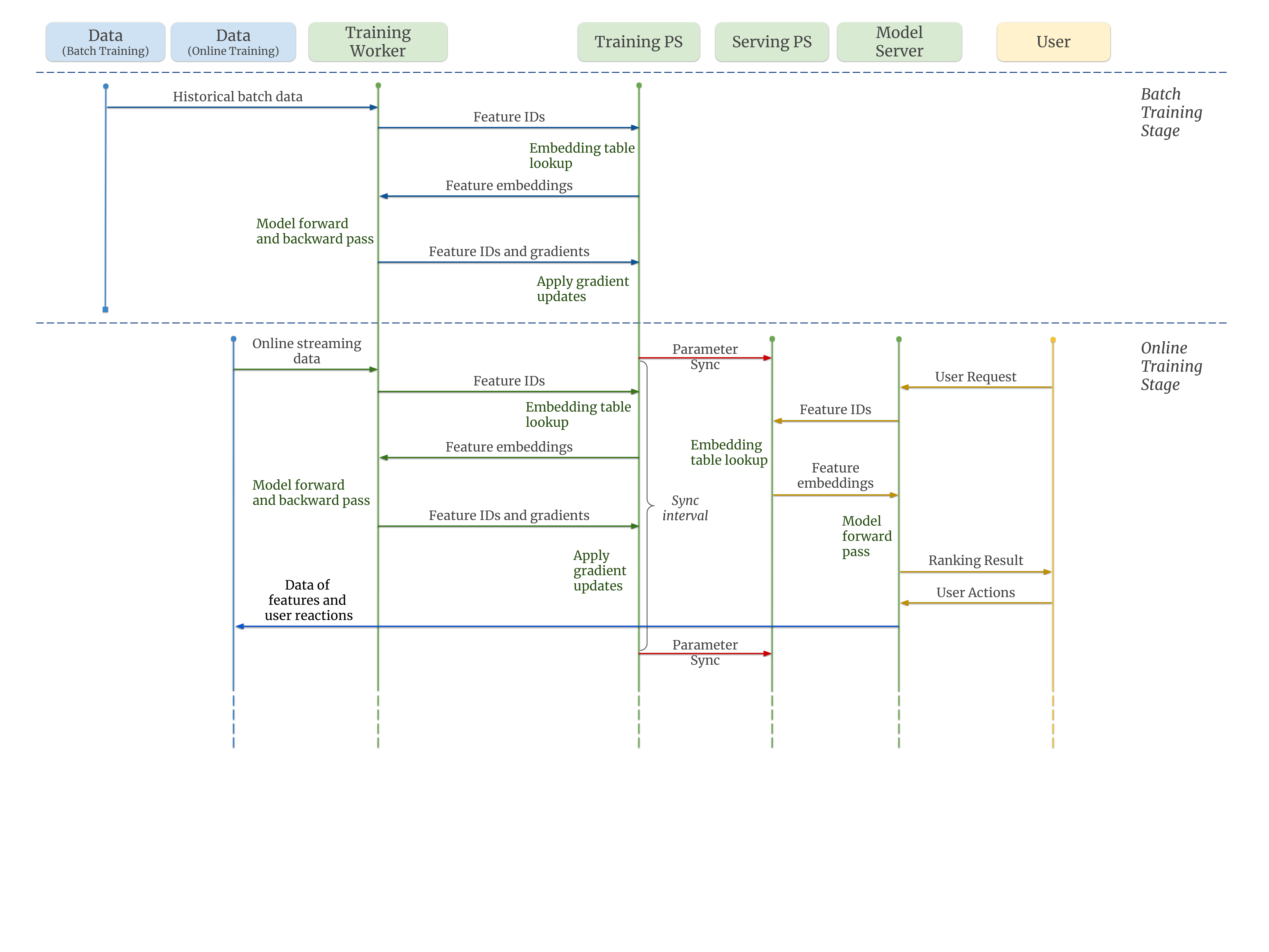}
  \caption{Monolith Online Training Architecture.}
  \label{fig:online-training}
\end{figure*}

\subsection{Sparsity and Dynamism}
The data for recommendation mostly contain sparse categorical features, some of which appear with low frequency.
The common practice of mapping them to a high-dimensional embedding space would give rise to a series of issues:
\begin{itemize}
    \item Unlike language models where number of word-pieces are limited, the amount of users and ranking items are orders of magnitude larger. Such an enormous embedding table would hardly fit into single host memory;
    \item Worse still, the size of embedding table is expected to grow over time as more users and items are admitted, while frameworks like \cite{tensorflow,pytorch} uses a fixed-size dense variables to represent embedding table.
\end{itemize}

In practice, many systems adopt low-collision hashing \cite{youtube-rec, instagram} as a way to reduce memory footprint and to allow growing of IDs. This relies on an over-idealistic assumption that IDs in the embedding table is distributed evenly in frequency, and collisions are harmless to the model quality. Unfortunately this is rarely true for a real-world recommendation system, where a small group of users or items have significantly more occurrences. With the organic growth of embedding table size, chances of hash key collision increases and lead to deterioration of model quality \cite{instagram}.
 
Therefore it is a natural demand for production-scale recommendation systems to have the capacity to capture as many features in its parameters, and also have the capability of elastically adjusting the number of users and items it tries to book-keep.

\subsection{Non-stationary Distribution}
Visual and linguistic patterns barely develop in a time scale of centuries, while the same user interested in one topic could shift their zeal every next minute. As a result, the underlying distribution of user data is non-stationary, a phenomenon commonly referred to as Concept Drift \cite{concept-drift}.

Intuitively, information from a more recent history can more effectively contribute to predicting the change in a user's behavior. To mitigate the effect of Concept Drift, serving models need to be updated from new user feedback as close to real-time as possible to reflect the latest interest of a user.
\hfill\break

\noindent In light of these distinction and in observation of issues that arises from our production, we designed \textbf{Monolith}, a large-scale recommendation system to address these pain-points. We did extensive experiments to verify and iterate our design in the production environment. Monolith is able to
\begin{enumerate}
    \item Provide full expressive power for sparse features by designing a collisionless hash table and a dynamic feature eviction mechanism;
    \item Loop serving feedback back to training in real-time with online training.
\end{enumerate}

Empowered by these architectural capacities, Monolith consistently outperforms systems that adopts hash-tricks with collisions with roughly similar memory usage, and achieves state-of-the-art online serving AUC without overly burdening our servers' computation power.

The rest of the paper is organized as follows. We first elaborate design details of how Monolith tackles existing challenge with collisionless hash table and realtime training in Section 2. Experiments and results will be presented in Section 3, along with production-tested conclusions and some discussion of trade-offs between time-sensitivity, reliability and model quality. Section 4 summarizes related work and compares them with Monolith. Section 5 concludes this work.

\begin{figure}[htbp!]
  \includegraphics[trim={0 9in 6in 0},clip,width=0.5\textwidth,height=0.5\textheight,keepaspectratio]{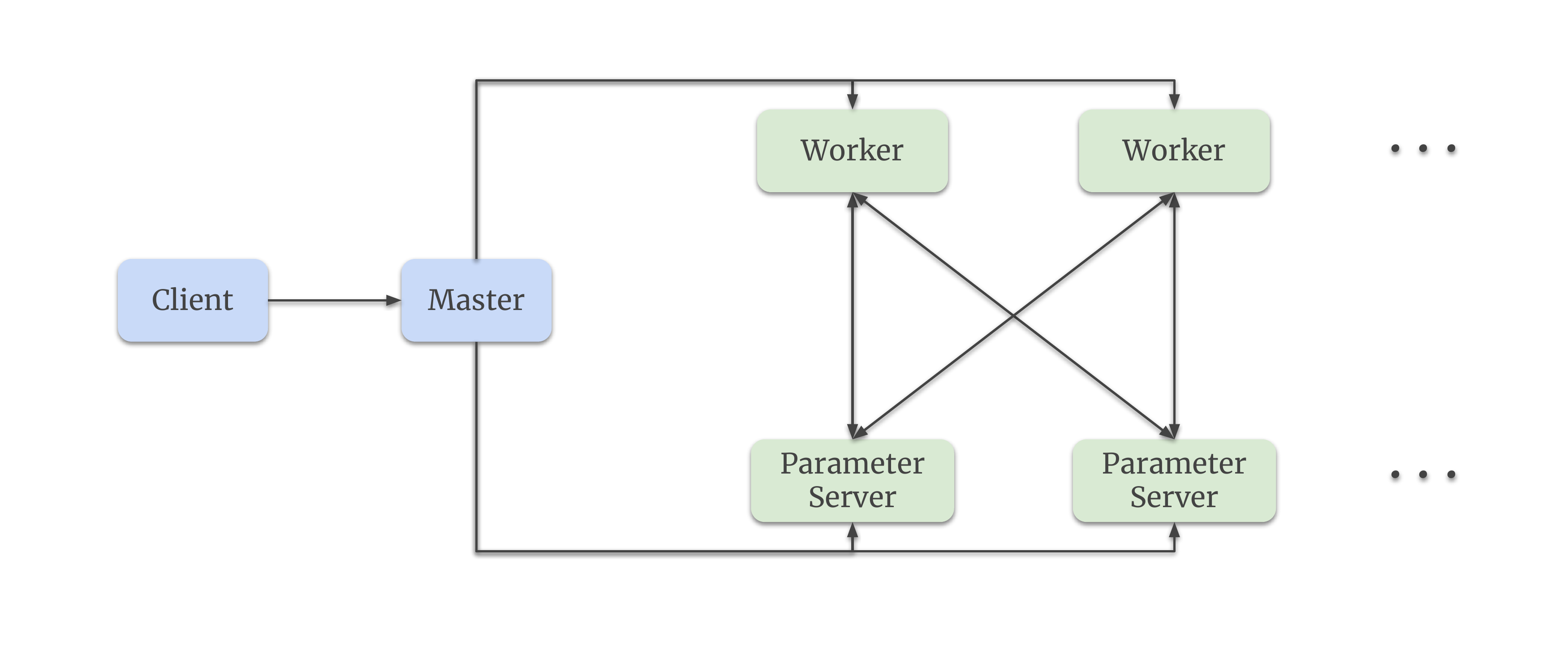}
  \caption{Worker-PS Architecture.}
  \label{fig:worker-ps}
\end{figure}

\begin{figure}[b!]
  \includegraphics[trim={0 3in 0.8in 0},clip,width=0.5\textwidth,height=0.5\textheight,keepaspectratio]{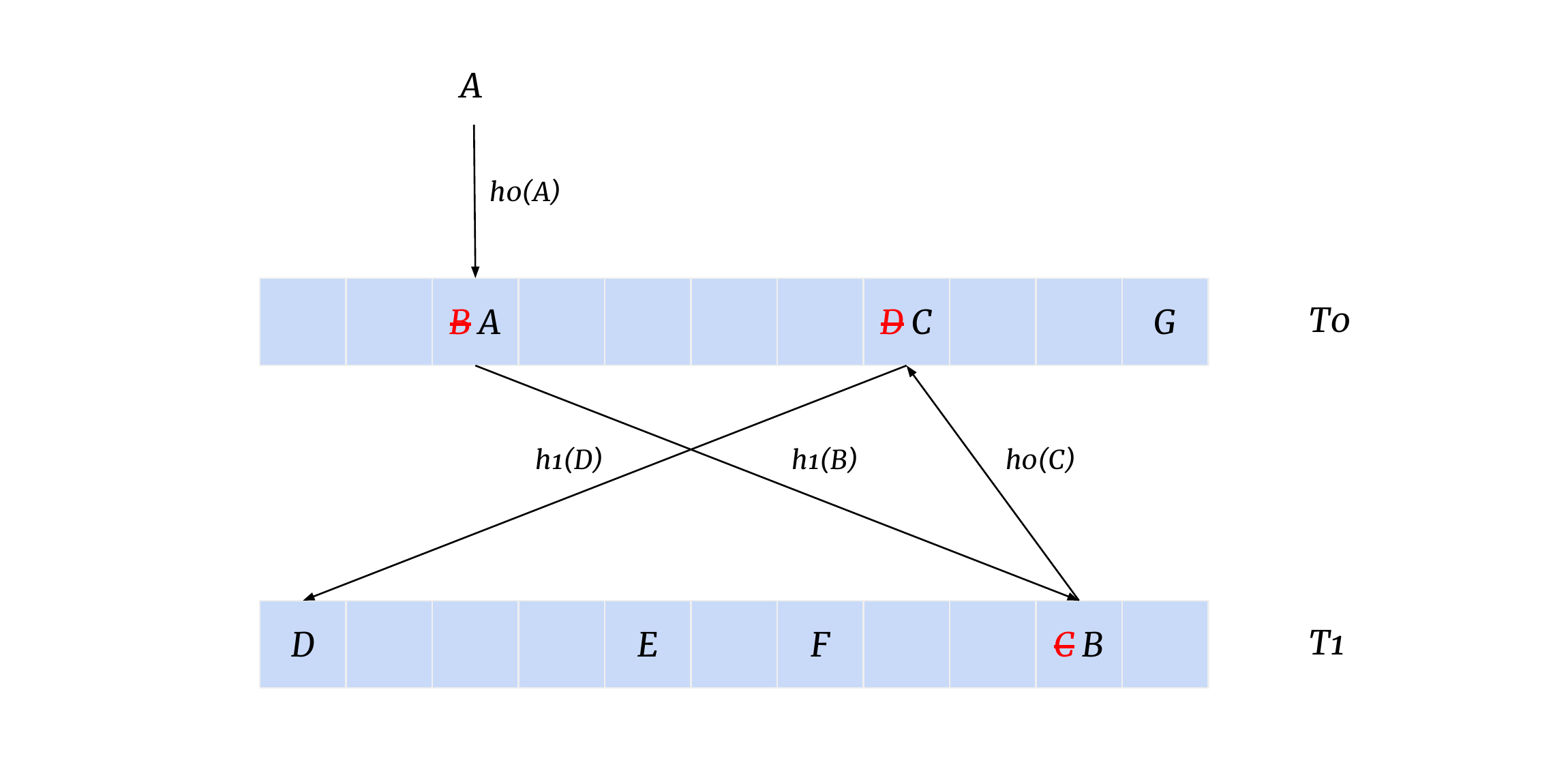}
  \caption{Cuckoo HashMap.}
  \label{fig:cuckoo}
\end{figure}

\section{Design}
\label{section:design}

The overall architecture of Monolith generally follows TensorFlow's distributed Worker-\textbf{P}arameter\textbf{S}erver setting (Figure \ref{fig:worker-ps}). In a Worker-PS architecture, machines are assigned different roles; Worker machines are responsible for performing computations as defined by the graph, and PS machines stores parameters and updates them according to gradients computed by Workers.

In recommendation models, parameters are categorized into two sets: dense and sparse. Dense parameters are weights/variables in a deep neural network, and sparse parameters refer to embedding tables that corresponds to sparse features. In our design, both dense and sparse parameters are part of TensorFlow Graph, and are stored on parameter servers.

Similar to TensorFlow's \textsf{Variable} for dense parameters, we designed a set of highly-efficient, collisionless, and flexible \textsf{HashTable} operations for sparse parameters. As an complement to TensorFlow's limitation that arises from separation of training and inference, Monolith's elastically scalable online training is designed to efficiently synchronize parameters from training-PS to online serving-PS within short intervals, with model robustness guarantee provided by fault tolerance mechanism.

\subsection{Hash Table}
\label{subsection:design/hash-table}
A first principle in our design of sparse parameter representation is to avoid cramping information from different IDs into the same fixed-size embedding. Simulating a dynamic size embedding table with an out-of-the-box TensorFlow \textsf{Variable} inevitably leads to ID collision, which exacerbates as new IDs arrive and table grows. Therefore instead of building upon \textsf{Variable}, we developed a new key-value \textsf{HashTable} for our sparse parameters.

Our \textsf{HashTable} utilizes Cuckoo Hashmap \cite{cuckoo} under the hood, which supports inserting new keys without colliding with existing ones. Cuckoo Hashing achieves worst-case $O(1)$ time complexity for lookups and deletions, and an expected amortized $O(1)$ time for insertions. As illustrated in Figure \ref{fig:cuckoo} it maintains two tables $T_0, T_1$ with different hash functions $h_0(x), h_1(x)$, and an element would be stored in either one of them. When trying to insert an element $A$ into $T_0$, it first attempts to place $A$ at $h_0(A)$; If $h_0(A)$ is occupied by another element $B$, it would evict $B$ from $T_0$ and try inserting $B$ into $T_1$ with the same logic. This process will be repeated until all elements stabilize, or rehash happens when insertion runs into a cycle.

\begin{figure*}[htbp!]
  \includegraphics[trim={0 7in 0 0},clip,width=\textwidth,height=\textheight,keepaspectratio]{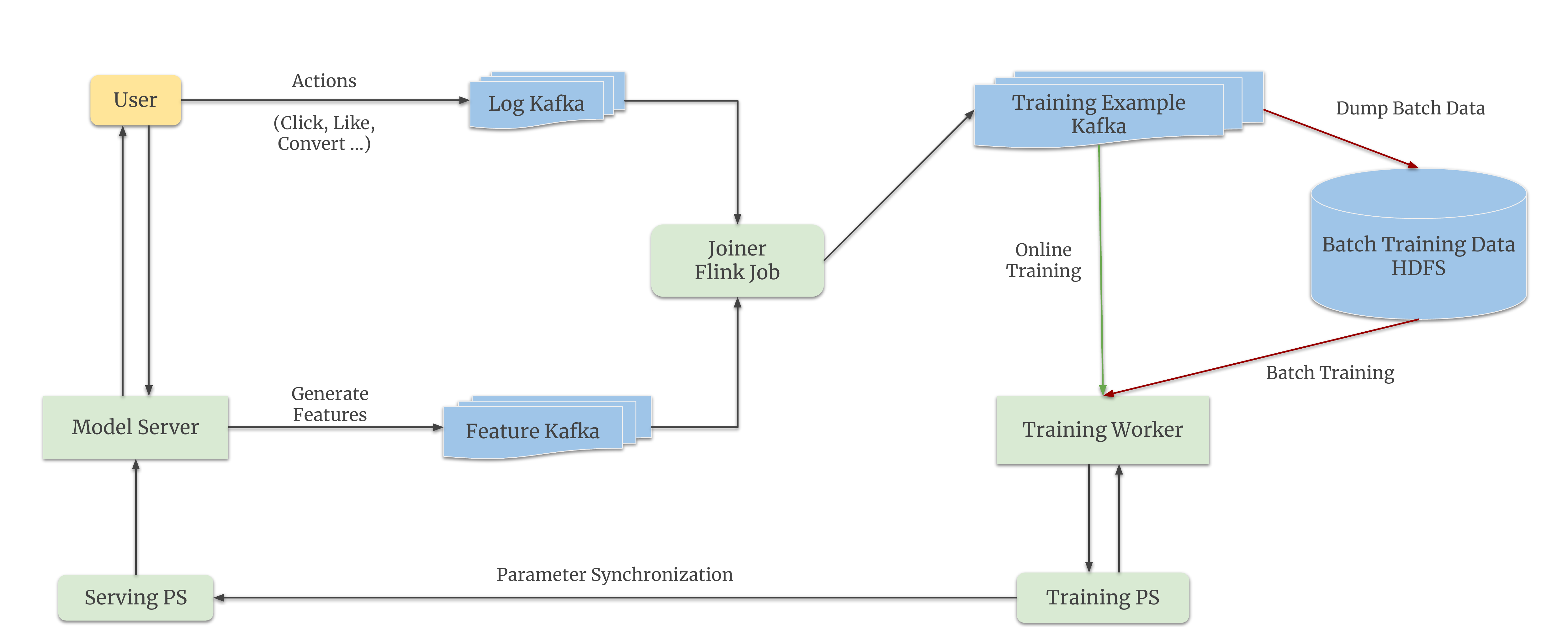}
  \caption{Streaming Engine.}
  \par \small \textit{The information feedback loop from [User $\rightarrow$ Model Server $\rightarrow$ Training Worker $\rightarrow$ Model Server $\rightarrow$ User] would spend a long time when taking the Batch Training path, while the Online Training will close the loop more instantly.}
  \label{fig:streaming}
\end{figure*}

Memory footprint reduction is also an important consideration in our design. A naive approach of inserting every new ID into the \textsf{HashTable} will deplete memory quickly. Observation of real production models lead to two conclusions:
\begin{enumerate}
    \item IDs that appears only a handful of times have limited contribution to improving model quality. An important observation is that IDs are long-tail distributed, where popular IDs may occur millions of times while the unpopular ones appear no more than ten times. Embeddings corresponding to these infrequent IDs are underfit due to lack of training data and the model will not be able to make a good estimation based on them. At the end of the day these IDs are not likely to affect the result, so model quality will not suffer from removal of these IDs with low occurrences;
    \item Stale IDs from a distant history seldom contribute to the current model as many of them are never visited. This could possibly due to a user that is no longer active, or a short-video that is out-of-date. Storing embeddings for these IDs could not help model in any way but to drain our PS memory in vain.
\end{enumerate}

Based on these observation, we designed several feature ID filtering heuristics for a more memory-efficient implementation of \textsf{HashTable}:
\begin{enumerate}
    \item IDs are filtered before they are admitted into embedding tables. We have two filtering methods: First we filter by their occurrences before they are inserted as keys, where the threshold of occurrences is a tunable hyperparameter that varies for each model; In addition we utilize a probabilistic filter which helps further reduce memory usage;
    \item IDs are timed and set to expire after being inactive for a predefined period of time. The expire time is also tunable for each embedding table to allow for distinguishing features with different sensitivity to historical information.
\end{enumerate}

In our implementation, \textsf{HashTable} is implemented as a TensorFlow resource operation. Similar to \textsf{Variable}, look-ups and updates are also implemented as native TensorFlow operations for easier integration and better compatibility.

\subsection{Online Training}

In Monolith, training is divided into two stages (Figure \ref{fig:online-training}):
\begin{enumerate}
    \item Batch training stage. This stage works as an ordinary TensorFlow training loop: In each training step, a training worker reads one mini-batch of training examples from the storage, requests parameters from PS, computes a forward and a backward pass, and finally push updated parameters to the training PS. Slightly different from other common deep learning tasks, we only train our dataset for one pass. Batch training is useful for training historical data when we modify our model architecture and retrain the model;
    \item Online training stage. After a model is deployed to online serving, the training does not stop but enters the online training stage. Instead of reading mini-batch examples from the storage, a training worker consumes realtime data on-the-fly and updates the training PS. The training PS periodically synchronizes its parameters to the serving PS, which will take effect on the user side immediately. This enables our model to interactively adapt itself according to a user's feedback in realtime.
\end{enumerate}

\begin{figure*}[htbp!]
  \includegraphics[trim={0 5in 0 1in},clip,width=\textwidth,height=\textheight,keepaspectratio]{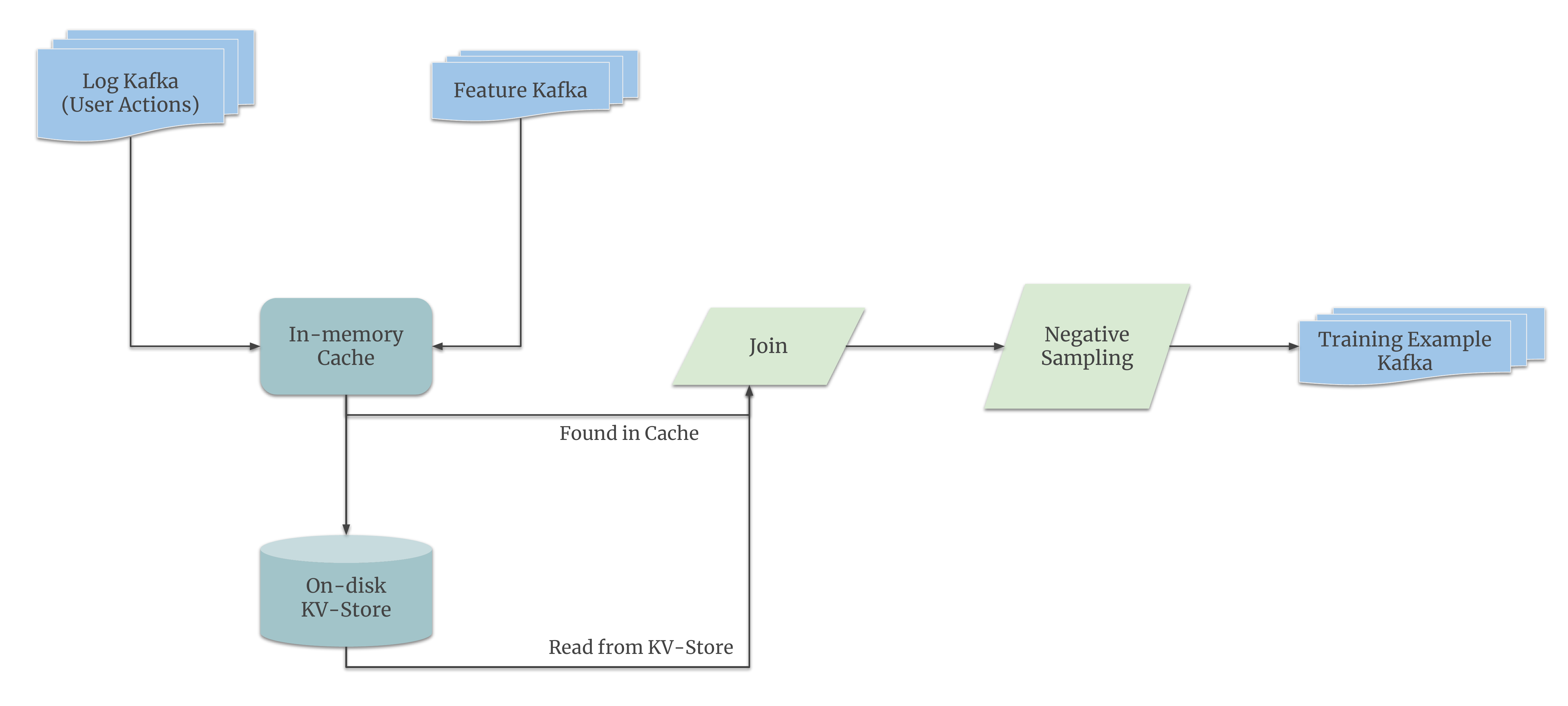}
  \caption{Online Joiner.}
  \label{fig:joiner}
\end{figure*}

\bigskip
\bigskip
\subsubsection{Streaming Engine}
Monolith is built with the capability of seamlessly switching between batch training and online training. This is enabled by 
our design of streaming engine as illustrated by Figure \ref{fig:streaming}.

In our design, we use one Kafka \cite{kafka} queue to log actions of users (E.g. Click on an item or like an item etc.) and another Kafka queue for features. At the core of the engine is a Flink \cite{flink} streaming job for online feature Joiner. The online joiner concatenates features with labels from user actions and produces training examples, which are then written to a Kafka queue. The queue for training examples is consumed by both online training and batch training:
\begin{itemize}
    \item For online training, the training worker directly reads data from the Kafka queue;
    \item For batch training, a data dumping job will first dump data to HDFS \cite{hadoop}; After data in HDFS accumulated to certain amount, training worker will retrieve data from HDFS and perform batch training.
\end{itemize}
Updated parameters in training PS will be pushed to serving PS according to the parameter synchronization schedule.

\subsubsection{Online Joiner}

In real-world applications, user actions log and features are streamed into the online joiner (Figure \ref{fig:joiner}) without guarantee in time order. Therefore we use a unique key for each request so that user action and features could correctly pair up.

The lag of user action could also be a problem. For example, a user may take a few days before they decide to buy an item they were presented days ago. This is a challenge for the joiner because if all features are kept in cache, it would simply not fit in memory. In our system, an on-disk key-value storage is utilized to store features that are waiting for over certain time period. When a user action log arrives, it first looks up the in-memory cache, and then looks up the key-value storage in case of a missing cache.

Another problem that arise in real-world application is that the distribution of negative and positive examples are highly uneven, where number of the former could be magnitudes of order higher than the latter. To prevent positive examples from being overwhelmed by negative ones, a common strategy is to do negative sampling. This would certainly change the underlying distribution of the trained model, tweaking it towards higher probability of making positive predictions. As a remedy, we apply log odds correction \cite{log-correction} during serving, making sure that the online model is an unbiased estimator of the original distribution.

\subsubsection{Parameter Synchronization}
\label{subsubsection:param-sync}
During online training, the Monolith training cluster keeps receiving data from the online serving module and updates parameters on the training PS. A crucial step to enable the online serving PS to benefit from these newly trained parameters is the synchronization of updated model parameters. In production environment, we are encountered by several challenges:
\begin{itemize}
    \item Models on the online serving PS must not stop serving when updating. Our models in production is usually several terabytes in size, and as a result replacing all parameters takes a while. It would be intolerable to stop an online PS from serving the model during the replacement process, and updates must be made on-the-fly;
    \item Transferring a multi-terabyte model of its entirety from training PS to the online serving PS would pose huge pressure to both the network bandwidth and memory on PS, since it requires doubled model size of memory to accept the newly arriving model.
\end{itemize}

For the online training to scale up to the size of our business scenario, we designed an incremental on-the-fly periodic parameter synchronization mechanism in Monolith based on several noticeable characteristic of our models:
\begin{enumerate}
    \item Sparse parameters are dominating the size of recommendation models;
    \item Given a short range of time window, only a small subset of IDs gets trained and their embeddings updated;
    \item Dense variables move much slower than sparse embeddings. This is because in momentum-based optimizers, the accumulation of momentum for dense variables is magnified by the gigantic size of recommendation training data, while only a few sparse embeddings receives updates in a single data batch.
\end{enumerate}
(1) and (2) allows us to exploit the sparse updates across all feature IDs. In Monolith, we maintain a hash set of touched keys, representing IDs whose embeddings get trained since the last parameter synchronization. We push the subset of sparse parameters whose keys are in the touched-keys set with a minute-level time interval from the training PS to the online serving PS. This relatively small pack of incremental parameter update is lightweight for network transmission and will not cause a sharp memory spike during the synchronization.

We also exploit (3) to further reduce network I/O and memory usage by setting a more aggressive sync schedule for sparse parameters, while updating dense parameters less frequently. This could render us a situation where the dense parameters we serve is a relatively stale version compared to sparse part. However, such inconsistency could be tolerated due to the reason mentioned in (3) as no conspicuous loss has been observed.

\subsection{Fault Tolerance}

As a system in production, Monolith is designed with the ability to recover a PS in case it fails. A common choice for fault tolerance is to snapshot the state of a model periodically, and recover from the latest snapshot when PS failure is detected. The choice of snapshot frequency has two major impacts:
\begin{enumerate}
    \item Model quality. Intuitively, model quality suffers less from loss of recent history with increased snapshot frequency.
    \item Computation overhead. Snapshotting a multi-terabyte model is not free. It incurs large chunks of memory copy and disk I/O.
\end{enumerate}

As a trade-off between model quality and computation overhead, Monolith snapshots all training PS every day. Though a PS will lose one day's worth of update in case of a failure, we discover that the performance degradation is tolerable through our experiments. We will analyze the effect of PS reliability in the next section.

\begin{figure}[htbp!]
  \includegraphics[trim={0 0 0 3in},clip,width=0.5\textwidth,height=0.5\textheight,keepaspectratio]{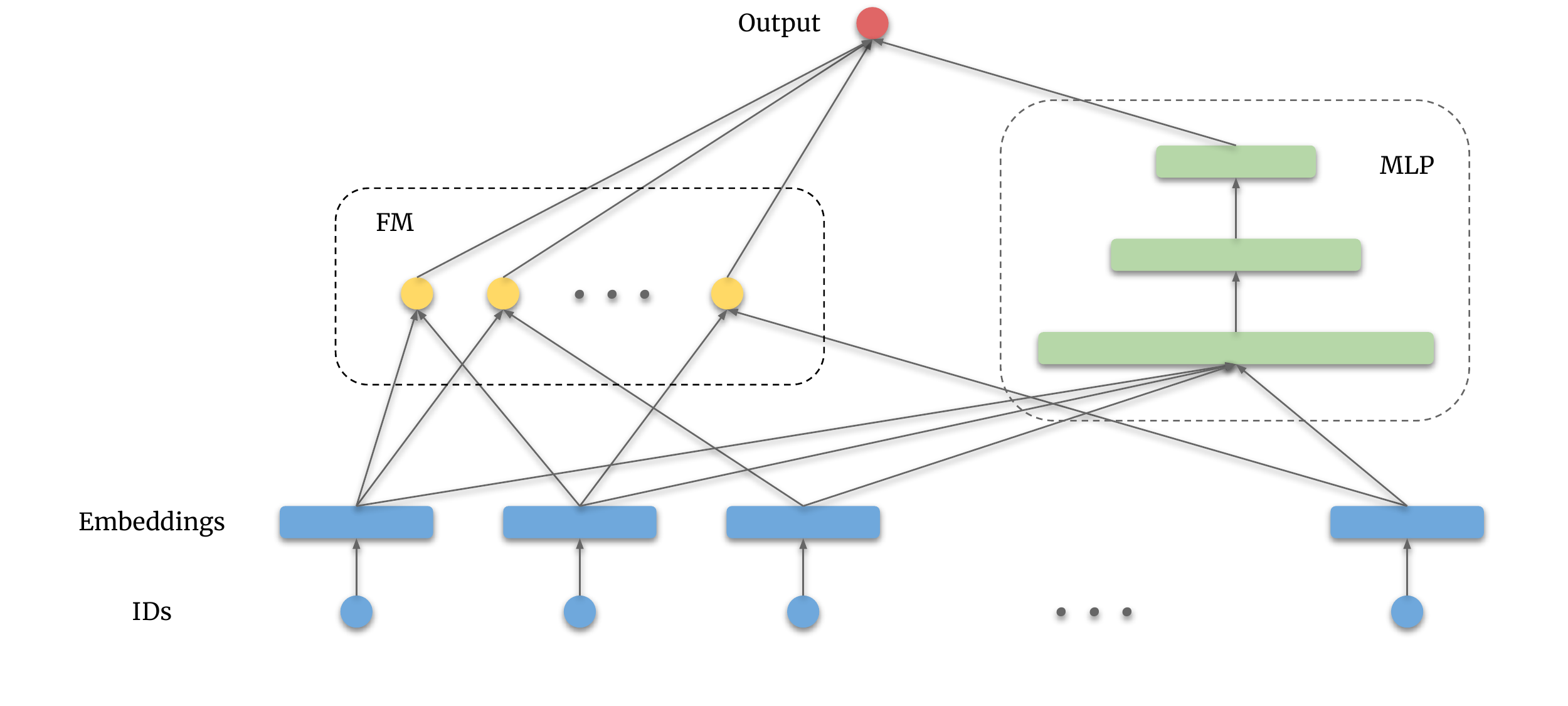}
  \caption{DeepFM model architecture.}
  \label{fig:deepfm}
\end{figure}

\section{Evaluation}

For a better understanding of benefits and trade-offs brought about by our proposed design, we conducted several experiments at production scale and A/B test with live serving traffic to evaluate and verify Monolith from different aspects. We aim to answer the following questions by our experiments:

\begin{enumerate}
    \item How much can we benefit from a collisionless \textsf{HashTable}?
    \item How important is realtime online training?
    \item Is Monolith's design of parameter synchronization robust enough in a large-scale production scenario?
\end{enumerate}

In this section, we first present our experimental settings and then discuss results and our findings in detail.

\subsection{Experimental Setup}

\subsubsection{Embedding Table}

As described in Section \ref{subsection:design/hash-table}, embedding tables in Monolith are implemented as collisionless \textsf{HashTable}s. To prove the necessity of avoiding collisions in embedding tables and to quantify gains from our collisionless implementation, we performed two groups of experiments on the Movielens dataset and on our internal production dataset respectively:

\begin{enumerate}
    \item \textbf{MovieLens} ml-25m dataset \cite{movielens}. This is a standard public dataset for movie ratings, containing 25 million ratings that involves approximately $162000$ users and $62000$ movies.
    \begin{itemize}
        \item \textit{Preprocessing of labels}. The original labels are ratings from 0.5 to 5.0, while in production our tasks are mostly receiving binary signals from users. To better simulate our production models, we convert scale labels to binary labels by treating scores $\ge 3.5$ as positive samples and the rest as negative samples.
        \item \textit{Model and metrics}. We implemented a standard DeepFM \cite{deepfm} model, a commonly used model architecture for recommendation problems. It consist of an FM component and a dense component (Figure \ref{fig:deepfm}). Predictions are evaluated by AUC \cite{auc} as this is the major measurement for real models.
        \item \textit{Embedding collisions}. This dataset contains approximately 160K user IDs and 60K movie IDs. To compare with the collisionless version of embedding table implementation, we performed another group of experiment where IDs are preprocessed with MD5 hashing and then mapped to a smaller ID space. As a result, some IDs will share their embedding with others. Table \ref{table:id-stat} shows detailed statistics of user and movie IDs before and after hashing.
        \begin{table}[h!]
        \centering
        \begin{tabular}{c  r  r} 
         \toprule
          & \textbf{User IDs} & \textbf{Movie IDs} \\
         \midrule
         \# Before Hashing & 162541 & 59047 \\
         \# After Hashing & 149970 & 57361 \\
         Collision rate & $7.73\%$ & $2.86\%$ \\
         \bottomrule
        \end{tabular}
        \par \medskip
        \caption{Statistics of IDs Before and After Hashing.}
        \label{table:id-stat}
        \end{table}
    \end{itemize}

    \item \textbf{Internal Recommendation dataset}.
    
    We also performed experiments on a recommendation model in production environment. This model generally follows a multi-tower architecture, with each tower responsible for learning to predict a specialized kind of user behavior.
    
    \begin{itemize}
        \item Each model has around 1000 embedding tables, and distribution of size of embedding tables are very uneven;
        \item The original ID space of embedding table was $2^{48}$. In our baseline, we applied a hashing trick by decomposing to curb the size of embedding table. To be more specific, we use two smaller embedding tables instead of a gigantic one to generate a unique embedding for each ID by vector combination:
        \begin{align*}
            ID_r &= ID\text{ \% }2^{24} \\
            ID_q &= ID \div 2^{24} \\
            E &= E_r + E_q,
        \end{align*}
        where $E_r, E_q$ are embeddings corresponding to $ID_r, ID_q$.
        This effectively reduces embedding table sizes from $2^{48}$ to $2^{25}$;
        \item This model is serving in real production, and the performance of this experiment is measured by online AUC with real serving traffic.
    \end{itemize}
    
\end{enumerate}

    \begin{figure*}[htbp!]
    \begin{minipage}{1.\linewidth}
      \begin{algorithm}[H]
        \caption{ Simulated Online Training.}
        \label{alg:online-training}
        \begin{algorithmic}[1]
        \STATE \textbf{Input:} $D^{batch}$  \tcc*[r]{Data for batch training.}
        \STATE \textbf{Input:} $D^{online}_{i=1 \cdots N}$  \tcc*[r]{Data for online training, split into $N$ shards. }
        \STATE $\theta_{train} \gets Train(D^{batch}, \theta_{train}) $    \tcc*[r]{Batch training.}
        \tcc*[h]{Online training.}
        \FOR{$i = 1 \cdots N$}     
            \STATE $\theta_{serve} \gets \theta_{train}$  \tcc*[r]{Sync training parameters to serving model.}
            \STATE $AUC_i = \texttt{Evaluate}(\theta_{serve}, D^{online}_i)$  \tcc*[r]{Evaluate online prediction on new data.}
            \STATE $\theta_{train} \gets Train(D^{online}_i, \theta_{train}) $  \tcc*[r]{Train with new data.}
        \ENDFOR
        \end{algorithmic}
        \end{algorithm}
    \end{minipage}
    \end{figure*}

\subsubsection{Online Training}

During online training, we update our online serving PS with the latest set of parameters with minute-level intervals. We designed two groups of experiments to verify model quality and system robustness.

\begin{enumerate}
    \item \textbf{Update frequency}. To investigate the necessity of minute-level update frequency, we conducted experiments that synchronize parameters from training model to prediction model with different intervals. 
    
    The dataset we use is the Criteo Display Ads Challenge dataset\footnote{\url{https://www.kaggle.com/competitions/criteo-display-ad-challenge/data}}, a large-scale standard dataset for benchmarking CTR models. It contains 7 days of chronologically ordered data recording features and click actions. For this experiment, we use a standard DeepFM \cite{deepfm} model as described in \ref{fig:deepfm}.
    
    To simulate online training, we did the following preprocessing for the dataset. We take 7 days of data from the dataset, and split it to two parts: 5 days of data for batch training, and 2 days for online training. We further split the 2 days of data into $N$ shards chronologically. Online training is simulated by algorithm \ref{alg:online-training}.
    
    As such, we simulate synchronizing trained parameters to online serving PS with an interval determined by number of data shards. We experimented with $N = 10, 50, 100$, which roughly correspond to update interval of $5 hr$, $1 hr$, and $30 min$.
    
    \item \textbf{Live experiment}. In addition, we also performed a live experiment with real serving traffic to further demonstrate the importance of online training in real-world application. This A/B experiment compares online training to batch training one one of our Ads model in production.

\end{enumerate}

\begin{figure}[htbp!]
    \centering
    \begin{minipage}{0.49\textwidth}
    \includegraphics[width=\textwidth,height=\textheight,keepaspectratio]{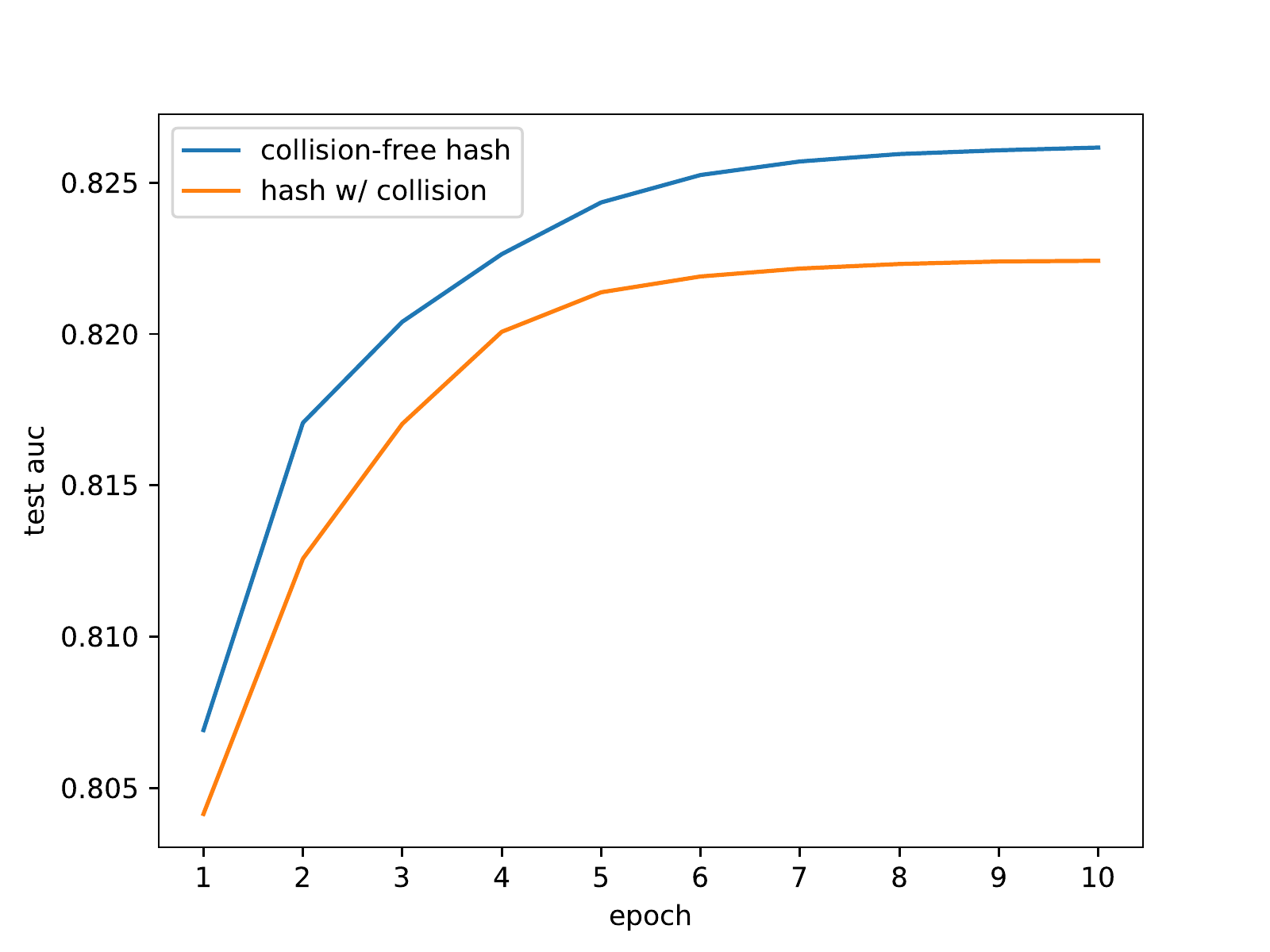}
    \caption{Effect of Embedding Collision On DeepFM, MovieLens}
    \label{fig:movielens-hash}
    \end{minipage}%
    \hfill
    \begin{minipage}{0.49\textwidth}
    \includegraphics[width=\textwidth,height=\textheight,keepaspectratio]{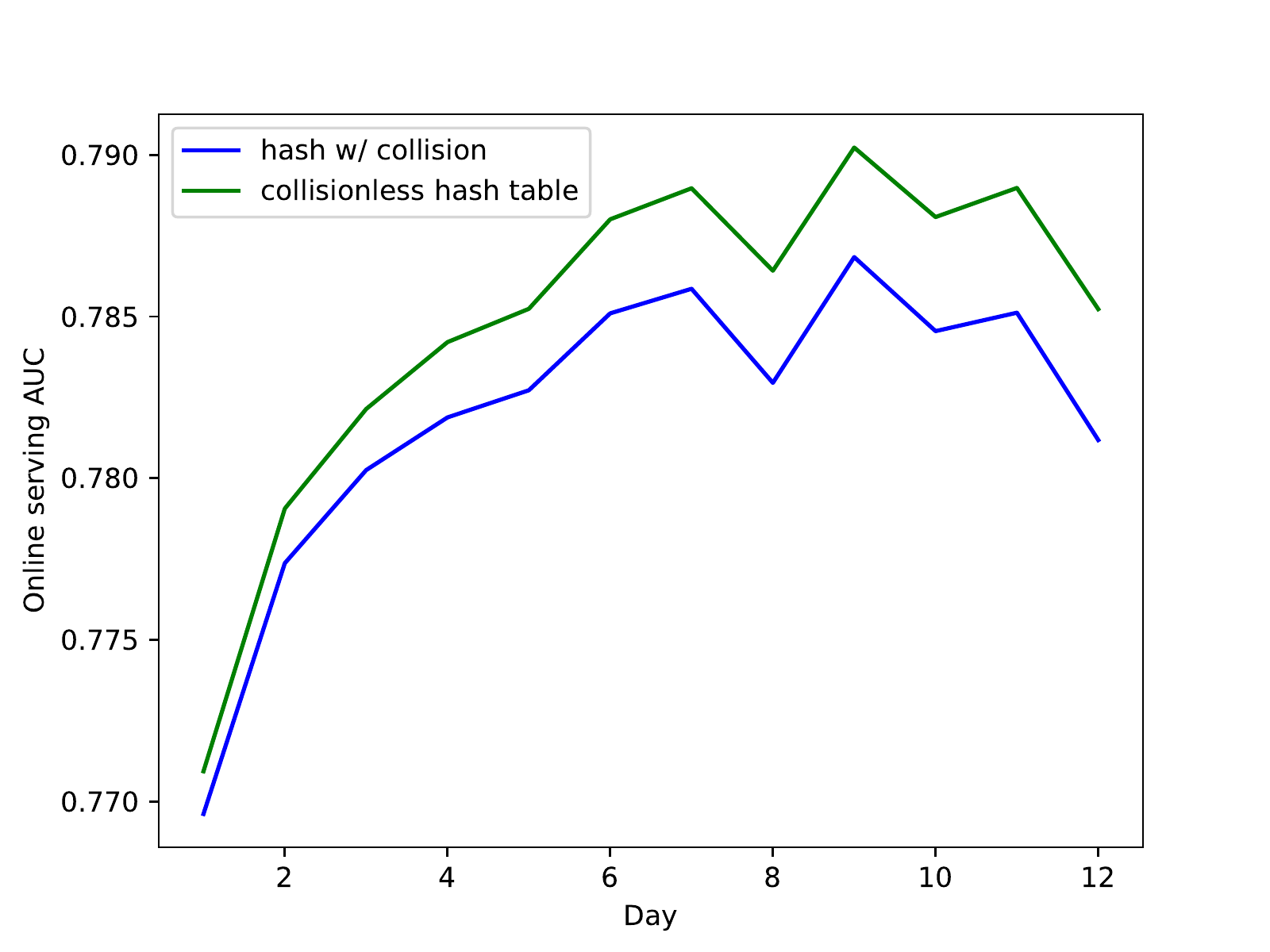}
    \caption{Effect of Embedding Collision On A Recommendation Model In Production}
    \par \small \textit{We measure performance of this recommendation model by online serving AUC, which is fluctuating across different days due to concept-drift.}
    \label{fig:cvr-hash}
    \end{minipage}%
\end{figure}

\subsection{Results and Analysis}

\subsubsection{The Effect of Embedding Collision}

Results from MovieLens dataset and the Internal recommedation dataset both show that embedding collisions will jeopardize model quality.
\begin{enumerate}
    \item Models with collisionless \textsf{HashTable} consistently outperforms those with collision. This conclusion holds true regardless of
    \begin{itemize}
        \item Increase of number of training epochs. As shown in Figure \ref{fig:movielens-hash}, the model with collisionless embedding table has higher AUC from the first epoch and converges at higher value;
        \item Change of distribution with passage of time due to Concept Drift. As shown in Figure \ref{fig:cvr-hash}, models with collisionless embedding table is also robust as time passes by and users/items context changes.
    \end{itemize}
    \item Data sparsity caused by collisionless embedding table will not lead to model overfitting. As shown in Figure \ref{fig:movielens-hash}, a model with collisionless embedding table does not overfit after it converges.
\end{enumerate}

    \begin{figure*}[htbp!]
        \centering
        \begin{subfigure}[t]{0.5\linewidth}
            \includegraphics[width=\linewidth]{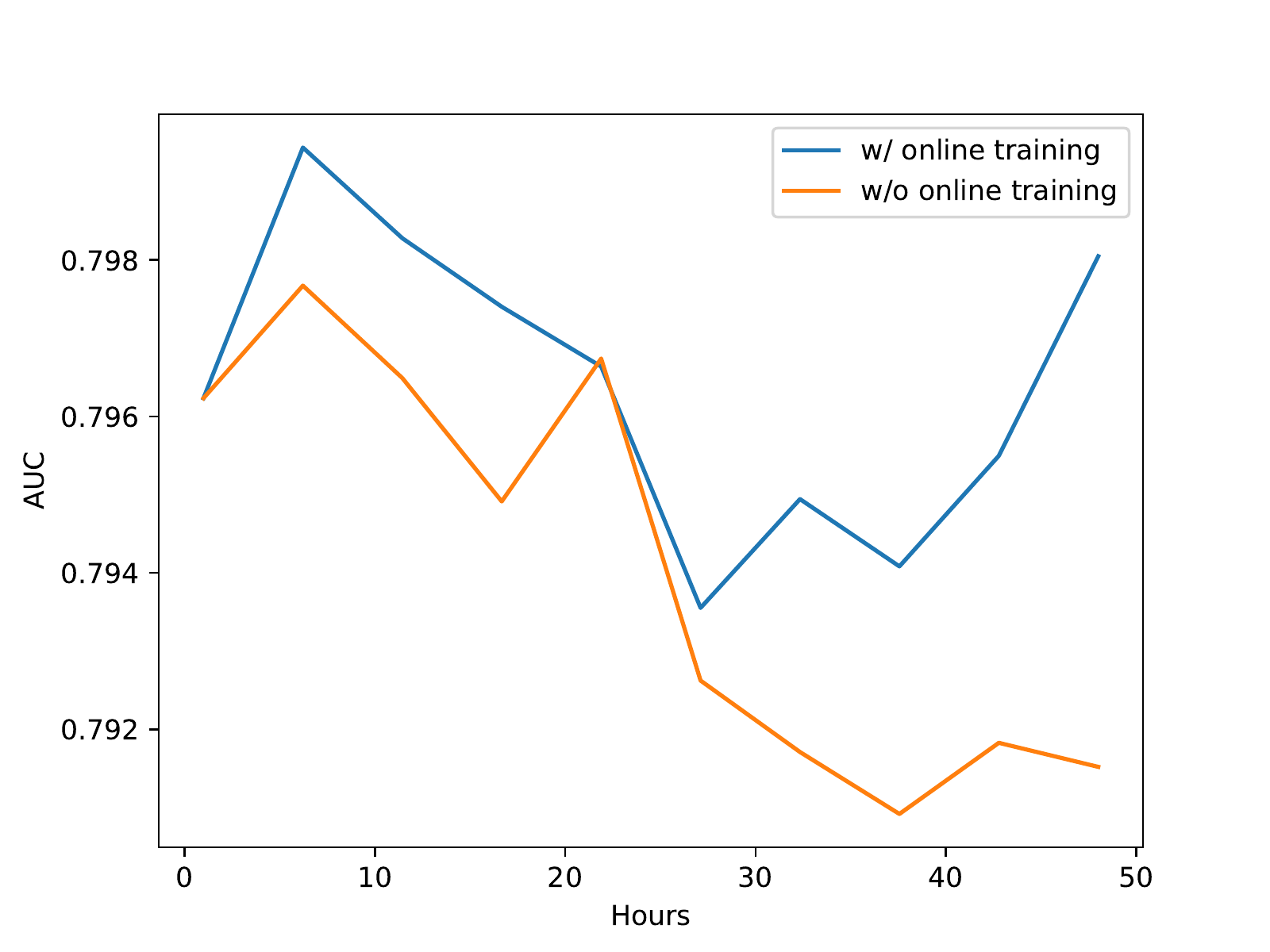}
            \caption{Online training with $5$ hrs sync interval}
            \label{fig:criteo-10}
        \end{subfigure}%
            \hfill%
        \begin{subfigure}[t]{0.5\linewidth}
            \includegraphics[width=\linewidth]{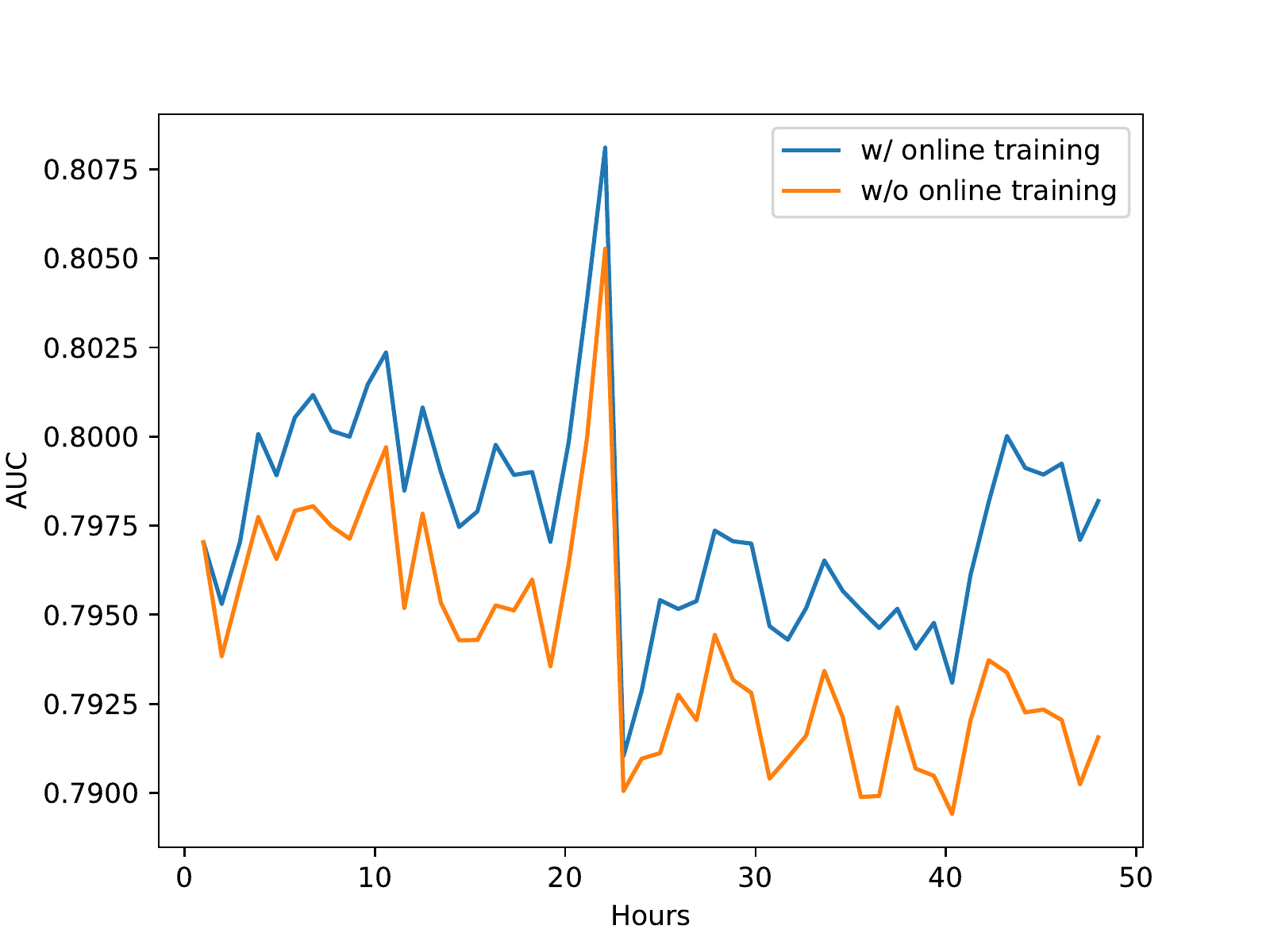}
            \caption{Online training with $1$ hr sync interval}
            \label{fig:criteo-50}
        \end{subfigure} 
        \begin{subfigure}[t]{0.5\linewidth}
            \includegraphics[width=\linewidth]{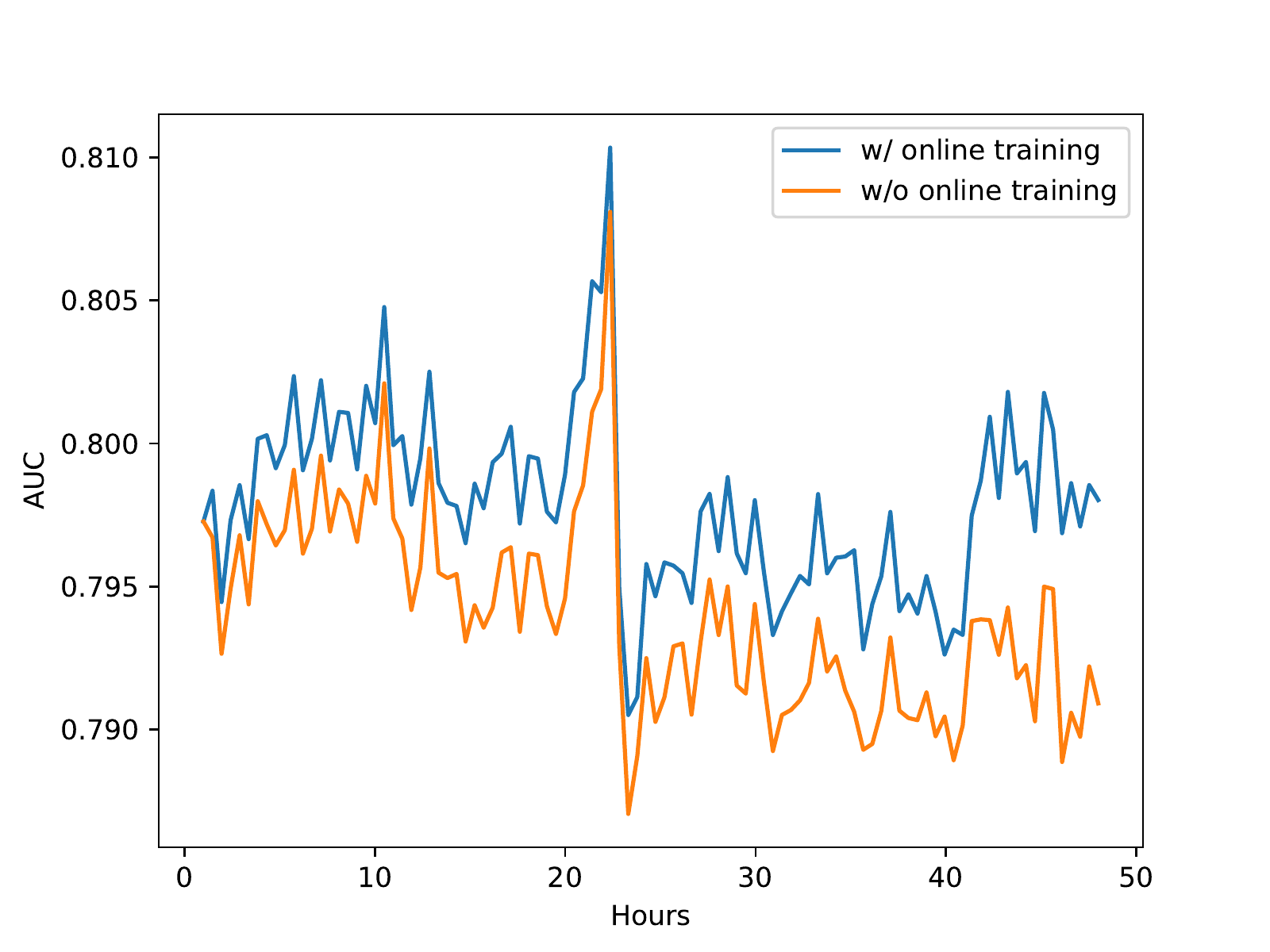}
            \caption{Online training with $30$ min sync interval}
            \label{fig:criteo-100}
        \end{subfigure}%
            \hfill%
        \caption{Online training v.s. Batch training on Criteo dataset.}
        \par \small \textit{Blue lines: AUC of models with online training; Yellow lines: AUC of batch training models evaluated against streaming data.}
        \label{fig:criteo}
    \end{figure*}

    \begin{figure}[htbp!]
      \includegraphics[width=0.5\textwidth,height=0.5\textheight,keepaspectratio]{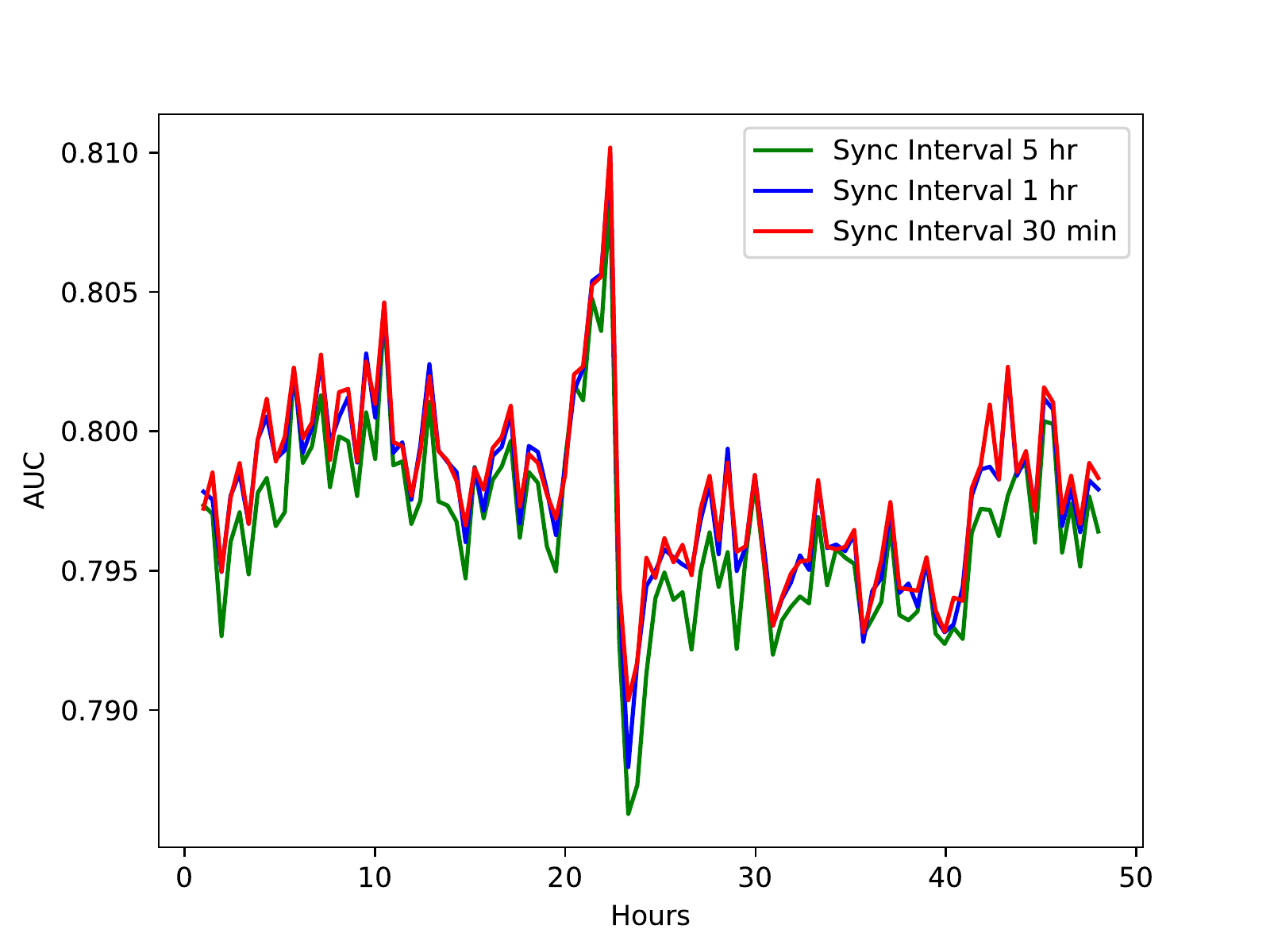}
      \caption{Comparison of different sync intervals for online training.}
      \label{fig:criteo-cmp}
    \end{figure}

\subsubsection{Online Training: Trading-off Reliability For Realtime} We discovered that a higher parameter synchronization frequency is always conducive to improving online serving AUC, and that online serving models are more tolerant with loss of a few shard of PS than we expect.

\begin{enumerate}
    \item \textbf{The Effect of Parameter Synchronization Frequency}. 
    
    In our online streaming training experiment (\ref{alg:online-training}) with Criteo Display Ads Challenge dataset, model quality consistently improves with the increase of parameter synchronization frequency, as is evident by comparison from two perspectives:
    \begin{itemize}
        \item Models with online training performs better than models without. Figure \ref{fig:criteo-10}, \ref{fig:criteo-50}, \ref{fig:criteo-100} compares AUC of online training models evaluated by the following shard of data versus batch training models evaluated by each shard of data;
        \item Models with smaller parameter synchronization interval performs better that those with larger interval. Figure \ref{fig:criteo-cmp} and Table \ref{table:criteo-avg} compares online serving AUC for models with sync interval of $5 hr$, $1 hr$, and $30 min$ respectively.
    \end{itemize}
    
    \begin{table}[htbp!]
        \centering
        \begin{tabular}{c c c} 
         \toprule
         \textbf{Sync Interval} & \textbf{Average AUC (online)} & \textbf{Average AUC (batch)} \\
         \midrule
         5 hr & $79.66 \pm 0.020$ & $79.42 \pm 0.026$ \\
         1 hr & $79.78 \pm 0.005$ & $79.44 \pm 0.030$ \\
         30 min & $\mathbf{79.80 \pm 0.008}$ & $79.43 \pm 0.025$ \\
         \bottomrule
        \end{tabular}
        \par \medskip
        \caption{Average AUC comparison for DeepFM model on Criteo dataset.}
        \label{table:criteo-avg}
    \end{table}

    \begin{table*}[htbp!]
        \centering
        \begin{tabular}{r | c c c c c c c} 
         \toprule
         \textbf{Day} & 1 & 2 & 3 & 4 & 5 & 6 & 7 \\
         \midrule
         \textbf{AUC Improvement \%} & 14.443 & 16.871 & 17.068 & 14.028 & 18.081 & 16.404 & 15.202 \\
         \bottomrule
        \end{tabular}
        \par \medskip
        \caption{Improvement of Online Training Over Batch Training from Live A/B Experiment on an Ads Model.}
        \label{table:prod-cmp}
    \end{table*}

    The live A/B experiment between online training and batch training on an Ads model in production also show that there is a significant bump in online serving AUC (Table \ref{table:prod-cmp}).
    
    Inspired by this observation, we synchronize sparse parameters to serving PS of our production models as frequent as possible (currently at minute-level), to the extent that the computation overhead and system reliability could endure. Recall that dense variables requires a less frequent update as discussed in \ref{subsubsection:param-sync}, we update them at day-level. By doing so, we can bring down our computation overhead to a very low level. Suppose 100,000 IDs gets updated in a minute, and the dimension of embedding is 1024, the total size of data need to be transferred is $4 KB \times 100,000 \approx 400 MB$ per minute. For dense parameters, since they are synchronized daily, we choose to schedule the synchronization when the traffic is lowest (e.g. midnight).

    \item \textbf{The Effect of PS reliability.}
    
    With a minute-level parameter synchronization, we initially expect a more frequent snapshot of training PS to match the realtime update. To our surprise, we enlarged the snapshot interval to 1 day and still observed nearly no loss of model quality.
    
    Finding the right trade-off between model quality and computation overhead is difficult for personalized ranking systems since users are extremely sensitive on recommendation quality. Traditionally, large-scale systems tend to set a frequent snapshot schedule for their models, which sacrifices computation resources in exchange for minimized loss in model quality.
    We also did quite some exploration in this regard and to our surprise, model quality is more robust than expected. With a $0.01\%$ failure rate of PS machine per day, we find a model from the previous day works embarrassingly well. This is explicable by the following calculation: Suppose a model's parameters are sharded across 1000 PS, and they snapshot every day. Given $0.01\%$ failure rate, one of them will go down every 10 days and we lose all updates on this PS for 1 day. Assuming a DAU of 15 Million and an even distribution of user IDs on each PS, we lose 1 day's feedback from 15000 users every 10 days. This is acceptable because (a) For sparse features which is user-specific, this is equivalent to losing a tiny fraction of $0.01\%$ DAU; (b) For dense variables, since they are updated slowly as we discussed in \ref{subsubsection:param-sync}, losing 1 day's update out of 1000 PS is negligible.

    Based on the above observation and calculation, we radically lowered our snapshot frequency and thereby saved quite a bit in computation overhead.

\end{enumerate}

\section{Related Work}
Ever since some earliest successful application of deep learning to industry-level recommendation systems \cite{youtube-rec,fb-dnn-rec}, researchers and engineers have been employing various techniques to ameliorate issues mentioned in Section \ref{section:introduction}.

To tackle the issue of sparse feature representation, \cite{youtube-rec, instagram} uses fixed-size embedding table with hash-trick. There are also attempts in improving hashing to reduce collision \cite{grubhub, instagram}. Other works directly utilize native key-value hash table to allow dynamic growth of table size \cite{kraken,xdl,meituan,aibox}. These implementations builds upon TensorFlow but relies either on specially designed software mechanism \cite{kraken,persia,meituan} or hardware \cite{aibox} to access and manage their hash-tables. Compared to these solutions, Monolith's hash-table is yet another native TensorFlow operation. It is developer friendly and has higher cross-platform interoperability, which is suitable for ToB scenarios. An organic and tight integration with TensorFlow also enables easier optimizations of computation performance.

Bridging the gap between training and serving and alleviation of Concept Drift \cite{concept-drift} is another topic of interest. To support online update and avoid memory issues, both \cite{xdl} and \cite{kraken} designed feature eviction mechanisms to flexibly adjust the size of embedding tables. Both \cite{xdl} and \cite{persia} support some form of online training, where learned parameters are synced to serving with a relatively short interval compared to traditional batch training, with fault tolerance mechanisms. Monolith took similar approach to elastically admit and evict features, while it has a more lightweight parameter synchronization mechanism to guarantee model quality.

\section{Conclusion}
In this work, we reviewed several most important challenges for industrial-level recommendation systems and present our system in production, Monolith, to address them and achieved best performance compared to existing solutions.

We proved that a collisionless embedding table is essential for model quality, and demonstrated that our implementation of Cuckoo HashMap based embedding table is both memory efficient and helpful for improving online serving metrics.

We also proved that realtime serving is crucial in recommendation systems, and that parameter synchronization interval should be as short as possible for an ultimate model performance. Our solution for online realtime serving in Monolith has a delicately designed parameter synchronization and a fault tolerance mechanism: In our parameter synchronization algorithm, we showed that consistency of version across different parts of parameters could be traded-off for reducing network bandwidth consumption; In fault tolerance design, we demonstrated that our strategy of trading-off PS reliability for realtime-ness is a robust solution.

To conclude, Monolith succeeded in providing a general solution for production scale recommendation systems.

%%
%% The acknowledgments section is defined using the "acks" environment
%% (and NOT an unnumbered section). This ensures the proper
%% identification of the section in the article metadata, and the
%% consistent spelling of the heading.
\begin{acks}
Hanzhi Zhou provided useful suggestions on revision of this paper.
\end{acks}

%%
%% The next two lines define the bibliography style to be used, and
%% the bibliography file.
\bibliographystyle{ACM-Reference-Format}
\bibliography{monolith}

%%
%% If your work has an appendix, this is the place to put it.
% \appendix

% \section{Research Methods}

% \subsection{Part One}

% Lorem ipsum dolor sit amet, consectetur adipiscing elit. Morbi
% malesuada, quam in pulvinar varius, metus nunc fermentum urna, id
% sollicitudin purus odio sit amet enim. Aliquam ullamcorper eu ipsum
% vel mollis. Curabitur quis dictum nisl. Phasellus vel semper risus, et
% lacinia dolor. Integer ultricies commodo sem nec semper.

% \subsection{Part Two}

% Etiam commodo feugiat nisl pulvinar pellentesque. Etiam auctor sodales
% ligula, non varius nibh pulvinar semper. Suspendisse nec lectus non
% ipsum convallis congue hendrerit vitae sapien. Donec at laoreet
% eros. Vivamus non purus placerat, scelerisque diam eu, cursus
% ante. Etiam aliquam tortor auctor efficitur mattis.

% \section{Online Resources}

% Nam id fermentum dui. Suspendisse sagittis tortor a nulla mollis, in
% pulvinar ex pretium. Sed interdum orci quis metus euismod, et sagittis
% enim maximus. Vestibulum gravida massa ut felis suscipit
% congue. Quisque mattis elit a risus ultrices commodo venenatis eget
% dui. Etiam sagittis eleifend elementum.

% Nam interdum magna at lectus dignissim, ac dignissim lorem
% rhoncus. Maecenas eu arcu ac neque placerat aliquam. Nunc pulvinar
% massa et mattis lacinia.

\end{document}